\documentclass[preprint,prc,aps]{revtex4}
\usepackage{latexsym}
\everymath={\displaystyle}
\usepackage{rotating}
\newcommand{\drawsquare}[2]{\hbox{%
\rule{#2pt}{#1pt}\hskip-#2pt
\rule{#1pt}{#2pt}\hskip-#1pt
\rule[#1pt]{#1pt}{#2pt}}\rule[#1pt]{#2pt}{#2pt}\hskip-#2pt
\rule{#2pt}{#1pt}}
\def\1{\mbox{l\hspace{-0.53em}1}}
\newcommand{\fr}{\frac}

\begin{document}
\title{Matrix Elements of SU(6) Generators for Baryons at Arbitrary $N_c$}

\author{N. Matagne\footnote{e-mail address: nmatagne@ulg.ac.be}}

\author{Fl. Stancu\footnote{ e-mail address: fstancu@ulg.ac.be
}}
\affiliation{University of Li\`ege, Institute of Physics B5, Sart Tilman,
B-4000 Li\`ege 1, Belgium}

\date{\today}

\begin{abstract}
We present explicit formulas for the matrix elements of the SU(6) generators
for totally symmetric spin-flavor states,
relevant for baryon spectroscopy in large $N_c$ QCD.
We rely on the interplay between two different methods
to calculate these matrix elements. As an outcome, general analytic
formulas of the corresponding SU(6)
isoscalar factors  are derived for arbitrary $N_c$. These results
can be used to study excited states of nonstrange and strange
baryons.
\end{abstract}

\maketitle

\section{Introduction}

The large $N_c$ limit of QCD  suggested by 't Hooft \cite{HOOFT} 
and the power counting rules of Witten \cite{WITTEN} lead to 
the powerful $1/N_c$ expansion method to study baryon spectroscopy. 
The method is based on
the result that the SU(2$N_f$) spin-flavor symmetry, 
where $N_f$ is the number of flavors, is exact in the
large $N_c$ limit of QCD \cite{DM}. 
For $N_c \rightarrow \infty $ the baryon masses are degenerate. 
For large $N_c$ the mass splitting starts at order $1/N_c$ for the ground state.
The method has been applied with great success
to the ground state baryons (N = 0 band), described
by the symmetric representation $\bf 56$ of SU(6)
\cite{DM,DJM94,DJM95,CGO94,Jenk1,JL95,DDJM96}.
Although the SU(6) symmetry is broken for excited states, 
it has been realized that the $1/N_c$ expansion can still be applied.

The excited states belonging to the $[{\bf 70},1^-]$ multiplet (N = 1 band)
\cite{CGKM,Goi97,PY1,PY2,CCGL,CaCa98,BCCG,SCHAT,Pirjol:2003ye,cohen1}
have been studied extensively in SU(4) ($N_f$ = 2).
The approach has been extended to  $N_f$ = 3 in Ref. \cite{SGS}
and it included first order in SU(3) symmetry breaking.  There are also
a few studies of the physically important multiplets 
belonging to  the N = 2 band. These are related to
$[{\bf 56'},0^+]$ \cite{CC00} in SU(4), to   $[{\bf 56},2^+]$ \cite{GSS} in
SU(6)  and to  $[{\bf 70},\ell^+]$ \cite{MS2} in SU(4).
The method of Ref. \cite{GSS}
has been applied to highly excited nonstrange and strange baryons
\cite{MS1} belonging to the  $[{\bf 56},4^+]$ multiplet (N = 4 band).
Configuration mixing have also been discussed \cite{GOITY05}.

In calculating the mass spectrum,
the general procedure is to split the baryon into an excited quark
and a core. The latter is in its ground state for the N = 1 band but
generally carries some excitation for N $>$ 1 (for example the
$[{\bf 70},\ell^+]$ multiplet \cite{MS2}). The excitation is implemented into the
orbital part of the wave function. The spin-flavor part
of the core wave function remains always symmetric.

The building blocks of the mass operator
describing baryons with $u$, $d$ and $s$ quarks are the excited quark operators
formed of the SO(3) generators $\ell^i_q$ and of the SU(6) generators
$s^i, t^a$ and $g^{ia}$ and the corresponding core operators 
$\ell^i_c$, $S^i_c, T^a_c$ and $G^{ia}_c$.
The matrix elements of the excited quark are straightforward, as being
single-particle operators. The matrix elements of the core operators 
$S^i_c, T^a_c$ are also simple to calculate, while those of $G^{ia}_c$
are more intricate. The purpose of this work is to derive explicit 
formulas for the matrix elements of these SU(6) generators, for
arbitrary $N_c$. So far,
only the SU(4) case was solved analytically \cite{PY1} for $G^{ia}_c$
and it was used as such in studies of
nonstrange baryons, as for example, in Ref. \cite{CCGL}.
Recently, explicit formulas for SU(3) Clebsch-Gordan coefficients, relevant 
for couplings of mesons to baryons at large $N_c$, have also been derived
\cite{CL}.

There are several ways to calculate the matrix elements of the SU(6) generators.
One is the standard group theory method. It is the way Hecht and Pang
\cite{HP} derived matrix elements of the SU(4) generators and it can straightforwardly 
be generalized
to SU(6). The difficulty in using this method is that it involves 
the knowledge of  isoscalar factors of SU(6). So far, the literature
provides a few examples of isoscalar factors:
${\bf 56} \times {\bf 35}  \rightarrow {\bf 56}$ \cite{CCM,SCHULKE},
${\bf 35} \times {\bf 35}  \rightarrow {\bf 35}$ \cite{SCHULKE} or
${\bf 35} \times {\bf 70} = {\bf 20}+{\bf 56}+2 \times {\bf 70}+{\bf 540}
+{\bf 560}+{\bf 1134}$ \cite{CC}
which can be applied to 
baryons composed of three quarks or to pentaquarks. 

Here we propose an alternative method, 
based on the decomposition of an SU(6) state into a product of
SU(3) and SU(2) states. It involves knowledge of isoscalar factors of
the permutation group S$_n$, with $n = N_c - 1$, for the core of a
baryon with an arbitrary number $N_c$ of quarks. As we shall see,
these isoscalar factors can easily be derived in various ways.

We recall that the group SU(6) has 35 generators ${S_i,T_a,G_{ia}}$
with $i = 1,2,3$ and $a = 1,2,...,8$, where $S_i$ are the generators 
of the spin subgroup SU(2) and $T_a$ the generators of the flavor 
subgroup SU(3). The group algebra is
\begin{eqnarray}\label{ALGEBRA}
[S_i,S_j] & = & i \varepsilon_{ijk} S_k,
\ \ \ [T_a,T_b]  =  i f_{abc} T_c, \ \ \ [S_i,T_a]  =  0,\nonumber \\
\lbrack S_i,G_{ia}\rbrack & = & i\varepsilon_{ijk}G_{ka}, \ \ \ [T_a,G_{ib}]=if_{abc}G_{ic}, \nonumber \\
\lbrack G_{ia},G_{jb}\rbrack & = & \fr{i}{4} \delta_{ij} f_{abc} T_c
+\fr{i}{2} \varepsilon_{ijk}\left(\fr{1}{3}\delta_{ab} S_k +d_{abc} G_{kc}\right),
\end{eqnarray}
by which the normalization of the generators is fixed.

The structure of the paper is as follows. In the next section 
we recall the standard group theory method to calculate the 
matrix elements of the generators of an unitary group.
In Sec. 3 we introduce the isoscalar factors of S$_n$ needed
to decompose a symmetric spin-flavor wave function of $N_c$ 
quarks into its spin and flavor parts. In Section 4 we derive 
matrix elements of the SU(6) generators between symmetric SU(6) states
by using the decomposition introduced in Sec. 3. 
In Sec. 5 we compare the results of the methods of Secs. 2 and 4. This allows us
to obtain explicit formulas for  the isoscalar factors of the SU(6) generators 
as a function of arbitrary $N_c$ and spin $S$. In Sec. 6, we return to the SU(4) case for completeness and consistency.
In the before last section we introduce the baryon mass operator
for which the above matrix elements are needed.  A summary is given in the
last section.

\section{SU(6) generators as tensor operators}

The SU(6) generators are the components of an irreducible tensor operator which transform
according to the adjoint representation $[21^4]$, equivalent to 
${\bf 35}$, in dimensional notation.   
The matrix elements of any irreducible tensor can be expressed in
terms of a generalized Wigner-Eckart theorem which is a factorization
theorem, involving the product between a reduced matrix element and 
a Clebsch-Gordan (CG) coefficient. 
The case SU(4) $\supset$ SU(2) $\times$ SU(2)
has been worked out by Hecht and Pang \cite{HP} and applied to
nuclear physics. 

Let us consider that the tensor operator $[21^4]$
acts on an SU(6) state of symmetry $[f]$. The symmetry of the final state,
denoted by $[f']$, labels one of the irreducible representations (irreps) appearing in the
Clebsch-Gordan  series
\begin{equation}\label{CGS}
[f] \times [21^4] = \sum_{[f']} m_{[f']}[f'],
\end{equation}
where $m_{[f']}$ denotes the multiplicity of the irrep $[f']$.
The multiplicity problem arises if $[f'] = [f]$.
An extra label $\rho$ is then necessary. 
It is not the case here in connection with SU(6). Indeed, if $N_c = 3$,
for the symmetric state  ${\bf 56}$, one has
${\bf 56} \times {\bf 35} \rightarrow {\bf 56}$ with multiplicity 1.
For arbitrary $N_c$ and $[f] = [N_c]$ the reduction (\ref{CGS}) in terms of Young diagrams reads
\begin{eqnarray}
\overbrace{\,\raisebox{-3.0pt}{\drawsquare{10.0}{0.4}}\hskip-0.4pt
        \raisebox{-3.0pt}{\drawsquare{10.0}{0.4}}\hskip-0.4pt
        \raisebox{-3.0pt}{\drawsquare{10.0}{0.4}}\, \raisebox{-1pt}{\mbox{$\cdots$}}\,
        \raisebox{-3.0pt}{\drawsquare{10.0}{0.4}}\,}^{N_c}  \times \
\raisebox{-23.0pt}{\drawsquare{10.0}{0.4}}\hskip-10.4pt
\raisebox{-13.0pt}{\drawsquare{10.0}{0.4}}\hskip-10.4pt
\raisebox{-3.0pt}{\drawsquare{10.0}{0.4}}\hskip-10.4pt
\raisebox{7.0pt}{\drawsquare{10.0}{0.4}}\hskip-10.4pt
        \raisebox{17pt}{\drawsquare{10.0}{0.4}}\hskip-0.4pt
        \raisebox{17pt}{\drawsquare{10.0}{0.4}} & = &
\overbrace{\,\raisebox{-3.0pt}{\drawsquare{10.0}{0.4}}\hskip-0.4pt
        \raisebox{-3.0pt}{\drawsquare{10.0}{0.4}}\, \raisebox{-1pt}{\mbox{$\cdots$}}\,
	\raisebox{-3.0pt}{\drawsquare{10.0}{0.4}}\hskip-0.4pt
        \raisebox{-3.0pt}{\drawsquare{10.0}{0.4}}\,}^{N_c} + \
\overbrace{\,\raisebox{-8.0pt}{\drawsquare{10.0}{0.4}}\hskip-10.4pt
        \raisebox{2.0pt}{\drawsquare{10.0}{0.4}}\hskip-0.4pt
        \raisebox{2.0pt}{\drawsquare{10.0}{0.4}}\, \raisebox{4pt}{\mbox{$\cdots$}} \,
        \raisebox{2.0pt}{\drawsquare{10.0}{0.4}}\,}^{N_c-1}  \nonumber \\
	& & + \ \overbrace{\raisebox{-23.0pt}{\drawsquare{10.0}{0.4}}\hskip-10.4pt
\raisebox{-13.0pt}{\drawsquare{10.0}{0.4}}\hskip-10.4pt
\raisebox{-3.0pt}{\drawsquare{10.0}{0.4}}\hskip-10.4pt
\raisebox{7.0pt}{\drawsquare{10.0}{0.4}}\hskip-0.4pt
\raisebox{7.0pt}{\drawsquare{10.0}{0.4}}\hskip-20.4pt
        \raisebox{17pt}{\drawsquare{10.0}{0.4}}\hskip-0.4pt
        \raisebox{17pt}{\drawsquare{10.0}{0.4}}\hskip-0.4pt
        \raisebox{17pt}{\drawsquare{10.0}{0.4}}\, \raisebox{19pt}{\mbox{$\cdots$}} \,
	\raisebox{17pt}{\drawsquare{10.0}{0.4}}\hskip-0.4pt
        \raisebox{17pt}{\drawsquare{10.0}{0.4}}\,}^{N_c+1} + \
\overbrace{\raisebox{-23.0pt}{\drawsquare{10.0}{0.4}}\hskip-10.4pt
\raisebox{-13.0pt}{\drawsquare{10.0}{0.4}}\hskip-10.4pt
\raisebox{-3.0pt}{\drawsquare{10.0}{0.4}}\hskip-10.4pt
\raisebox{7.0pt}{\drawsquare{10.0}{0.4}}\hskip-10.4pt
        \raisebox{17pt}{\drawsquare{10.0}{0.4}}\hskip-0.4pt
        \raisebox{17pt}{\drawsquare{10.0}{0.4}}\hskip-0.4pt
        \raisebox{17pt}{\drawsquare{10.0}{0.4}}\, \raisebox{19pt}{\mbox{$\cdots$}} \,
	\raisebox{17pt}{\drawsquare{10.0}{0.4}}\hskip-0.4pt
        \raisebox{17pt}{\drawsquare{10.0}{0.4}}\hskip-0.4pt
        \raisebox{17pt}{\drawsquare{10.0}{0.4}}}^{N_c+2},
\end{eqnarray}
which gives $m_{[f']} = 1$ for all terms, including the case $[f'] = [f]$.
But the multiplicity problem arises at the level of the subgroup
SU(3). Introducing the  SU(3) $\times$ SU(2) content of the ${\bf 56}$ and
${\bf 35}$ irreps into their direct product one finds that the product
${\bf 8} \times {\bf 8}$ appears twice. For arbitrary $N_c$, this product is given by
\begin{eqnarray}
\overbrace{\raisebox{-8.0pt}{\drawsquare{10.0}{0.4}}\hskip-10.4pt
        \raisebox{2pt}{\drawsquare{10.0}{0.4}}\hskip-0.4pt
\raisebox{-8.0pt}{\drawsquare{10.0}{0.4}}\hskip-10.4pt
        \raisebox{2pt}{\drawsquare{10.0}{0.4}}\hskip-0.4pt
\raisebox{-8.0pt}{\drawsquare{10.0}{0.4}}\hskip-10.4pt
        \raisebox{2pt}{\drawsquare{10.0}{0.4}}
\, \raisebox{0pt}{\mbox{$\cdots$}} \,
\raisebox{-8.0pt}{\drawsquare{10.0}{0.4}}\hskip-10.4pt
        \raisebox{2pt}{\drawsquare{10.0}{0.4}}\hskip-0.4pt
        \raisebox{2pt}{\drawsquare{10.0}{0.4}}\hskip-0.4pt}^{\frac{N_c+1}{2}}\, \times \
\raisebox{-8.0pt}{\drawsquare{10.0}{0.4}}\hskip-10.4pt
        \raisebox{2pt}{\drawsquare{10.0}{0.4}}\hskip-0.4pt
        \raisebox{2pt}{\drawsquare{10.0}{0.4}}\hskip-0.4pt \, & = &
\overbrace{\raisebox{-8.0pt}{\drawsquare{10.0}{0.4}}\hskip-10.4pt
        \raisebox{2pt}{\drawsquare{10.0}{0.4}}\,
\raisebox{0pt}{\mbox{$\cdots$}} \,
\raisebox{-8.0pt}{\drawsquare{10.0}{0.4}}\hskip-10.4pt
        \raisebox{2pt}{\drawsquare{10.0}{0.4}}\hskip-0.4pt
        \raisebox{2pt}{\drawsquare{10.0}{0.4}}\hskip-0.4pt
        \raisebox{2pt}{\drawsquare{10.0}{0.4}}\hskip-0.4pt}^{\frac{N_c-1}{2}}\,
	+ \
\overbrace{\raisebox{-8.0pt}{\drawsquare{10.0}{0.4}}\hskip-10.4pt
        \raisebox{2pt}{\drawsquare{10.0}{0.4}}\hskip-0.4pt
\raisebox{-8.0pt}{\drawsquare{10.0}{0.4}}\hskip-10.4pt
        \raisebox{2pt}{\drawsquare{10.0}{0.4}}\hskip-0.4pt
\raisebox{-8.0pt}{\drawsquare{10.0}{0.4}}\hskip-10.4pt
        \raisebox{2pt}{\drawsquare{10.0}{0.4}}\,
\raisebox{0pt}{\mbox{$\cdots$}} \,
\raisebox{-8.0pt}{\drawsquare{10.0}{0.4}}\hskip-10.4pt
        \raisebox{2pt}{\drawsquare{10.0}{0.4}}\hskip-0.4pt
\raisebox{-8.0pt}{\drawsquare{10.0}{0.4}}\hskip-10.4pt
        \raisebox{2pt}{\drawsquare{10.0}{0.4}}\hskip-0.4pt
        \raisebox{2pt}{\drawsquare{10.0}{0.4}}\hskip-0.4pt
        \raisebox{2pt}{\drawsquare{10.0}{0.4}}\hskip-0.4pt}^{\frac{N_c+5}{2}}\,  \nonumber \\
 & & + \overbrace{\raisebox{-8.0pt}{\drawsquare{10.0}{0.4}}\hskip-10.4pt
        \raisebox{2pt}{\drawsquare{10.0}{0.4}}\,
\raisebox{0pt}{\mbox{$\cdots$}} \,
\raisebox{-8.0pt}{\drawsquare{10.0}{0.4}}\hskip-10.4pt
        \raisebox{2pt}{\drawsquare{10.0}{0.4}}\hskip-0.4pt
\raisebox{-8.0pt}{\drawsquare{10.0}{0.4}}\hskip-10.4pt
        \raisebox{2pt}{\drawsquare{10.0}{0.4}}}^{\frac{N_c-3}{2}}\, + \
\overbrace{\raisebox{-8.0pt}{\drawsquare{10.0}{0.4}}\hskip-10.4pt      \raisebox{2pt}{\drawsquare{10.0}{0.4}}\hskip-0.4pt
\raisebox{-8.0pt}{\drawsquare{10.0}{0.4}}\hskip-10.4pt
        \raisebox{2pt}{\drawsquare{10.0}{0.4}}\hskip-0.4pt
\raisebox{-8.0pt}{\drawsquare{10.0}{0.4}}\hskip-10.4pt
        \raisebox{2pt}{\drawsquare{10.0}{0.4}}\,
\raisebox{0pt}{\mbox{$\cdots$}} \,
\raisebox{-8.0pt}{\drawsquare{10.0}{0.4}}\hskip-10.4pt
        \raisebox{2pt}{\drawsquare{10.0}{0.4}}\hskip-0.4pt
\raisebox{-8.0pt}{\drawsquare{10.0}{0.4}}\hskip-10.4pt
        \raisebox{2pt}{\drawsquare{10.0}{0.4}}\hskip-0.4pt
\raisebox{-8.0pt}{\drawsquare{10.0}{0.4}}\hskip-10.4pt
        \raisebox{2pt}{\drawsquare{10.0}{0.4}}}^{\frac{N_c+3}{2}}\,  \nonumber \\
& & + \left(\,\,
\overbrace{\raisebox{-8.0pt}{\drawsquare{10.0}{0.4}}\hskip-10.4pt
        \raisebox{2pt}{\drawsquare{10.0}{0.4}}\hskip-0.4pt
\raisebox{-8.0pt}{\drawsquare{10.0}{0.4}}\hskip-10.4pt
        \raisebox{2pt}{\drawsquare{10.0}{0.4}}\,
\raisebox{0pt}{\mbox{$\cdots$}} \,
\raisebox{-8.0pt}{\drawsquare{10.0}{0.4}}\hskip-10.4pt
        \raisebox{2pt}{\drawsquare{10.0}{0.4}}\hskip-0.4pt
\raisebox{-8.0pt}{\drawsquare{10.0}{0.4}}\hskip-10.4pt
        \raisebox{2pt}{\drawsquare{10.0}{0.4}}\hskip-0.4pt
        \raisebox{2pt}{\drawsquare{10.0}{0.4}}\hskip-0.4pt}^{\frac{N_c+1}{2}}\,\right)_1\, + \
 \left(\,\,\overbrace{\raisebox{-8.0pt}{\drawsquare{10.0}{0.4}}\hskip-10.4pt
        \raisebox{2pt}{\drawsquare{10.0}{0.4}}\hskip-0.4pt
\raisebox{-8.0pt}{\drawsquare{10.0}{0.4}}\hskip-10.4pt
        \raisebox{2pt}{\drawsquare{10.0}{0.4}}\,
\raisebox{0pt}{\mbox{$\cdots$}} \,
\raisebox{-8.0pt}{\drawsquare{10.0}{0.4}}\hskip-10.4pt
        \raisebox{2pt}{\drawsquare{10.0}{0.4}}\hskip-0.4pt
\raisebox{-8.0pt}{\drawsquare{10.0}{0.4}}\hskip-10.4pt
        \raisebox{2pt}{\drawsquare{10.0}{0.4}}\hskip-0.4pt
        \raisebox{2pt}{\drawsquare{10.0}{0.4}}\hskip-0.4pt}^{\frac{N_c+1}{2}}\,\right)_2\, \nonumber \\
& & + \
\overbrace{\raisebox{-8.0pt}{\drawsquare{10.0}{0.4}}\hskip-10.4pt
        \raisebox{2pt}{\drawsquare{10.0}{0.4}}\hskip-0.4pt
\raisebox{-8.0pt}{\drawsquare{10.0}{0.4}}\hskip-10.4pt
        \raisebox{2pt}{\drawsquare{10.0}{0.4}} \,
\raisebox{0pt}{\mbox{$\cdots$}} \,
\raisebox{-8.0pt}{\drawsquare{10.0}{0.4}}\hskip-10.4pt
        \raisebox{2pt}{\drawsquare{10.0}{0.4}}\hskip-0.4pt
        \raisebox{2pt}{\drawsquare{10.0}{0.4}}\hskip-0.4pt
        \raisebox{2pt}{\drawsquare{10.0}{0.4}}\hskip-0.4pt
        \raisebox{2pt}{\drawsquare{10.0}{0.4}}}^{\frac{N_c+3}{2}}\,.
\end{eqnarray}
In the following, $\rho$ is used to distinguish between
various  ${\bf 8} \times {\bf 8}$ products. Then this label is carried
over by the isoscalar factors of SU(6) (see below).
In particle physics one uses the label  $s$ for the symmetric
and $a$ for the antisymmetric product (see \emph{e.g.} \cite{CCM}).
Here we shall use the notation $\rho$ of Hecht \cite{HECHT}. For $N_c$ = 3
the relation to other labels is: $\rho$ = 1 corresponds to $(8 \times 8)_a$
or to $(8 \times 8)_2$ of De Swart \cite{DESWART}; $\rho$ = 2 corresponds to
$(8 \times 8)_s$ or to  $(8 \times 8)_1$ of De Swart. Throughout the paper,
we shall use the SU(3) notations and phase conventions of Hecht \cite{HECHT}.
Accordingly an irreducible representation of SU(3) carries the label $(\lambda \mu)$,
 introduced by Elliott \cite{ELLIOTT}, who applied SU(3) for the first
time in physics, to describe rotational bands of deformed nuclei \cite{book}. In particle physics the corresponding notation is $(p,q)$.
By analogy to SU(4)  \cite{HP} one can write
the  matrix elements of the SU(6) generators $E_{ia}$ as
\begin{eqnarray}\label{GEN}
\langle [N_c](\lambda' \mu') Y' I' I'_3 S' S'_3 | E_{ia} |
[N_c](\lambda \mu) Y I I_3 S S_3 \rangle = \sqrt{C(\mathrm{SU(6)})}  
  \left(\begin{array}{cc|c}
	S   &    S^i   & S'   \\
	S_3  &   S^i_3   & S'_3
  \end{array}\right)
     \left(\begin{array}{cc|c}
	I   &   I^a   & I'   \\
	I_3 &   I^a_{3}   & I'_3
   \end{array}\right)  \nonumber \\
 \times       \sum_{\rho = 1,2}
 \left(\begin{array}{cc||c}
	(\lambda \mu)    &  (\lambda^a\mu^a)   &   (\lambda' \mu')\\
	Y I   &  Y^a I^a  &  Y' I'
      \end{array}\right)_{\rho}
\left(\begin{array}{cc||c}
	[N_c]    &  [21^4]   & [N_c]   \\
	(\lambda \mu) S  &  (\lambda^a\mu^a) S^i  &  (\lambda' \mu') S'
      \end{array}\right)_{\rho} , 
   \end{eqnarray}
where $C(\mathrm{SU(6)}) = 5[N_c(N_c+6)]/12$ is the Casimir operator of SU(6), followed by Clebsch-Gordan
coefficients of SU(2)-spin and SU(2)-isospin. The sum over $\rho$ is over
terms containing products of isoscalar factors of SU(3) and SU(6) respectively.
We introduce $T_a$ as an SU(3) irreducible tensor operator of components
$T^{(11)}_{Y^aI^a}$. It is a scalar in
SU(2) so that the index $i$ is no more necessary. The generators $S_i$
form a rank 1 tensor in SU(2) which is a scalar in SU(3), so the index $i$ suffices.
Although we use the same symbol for the operator $S_i$ and its quantum numbers we hope that no confusion is created.
The relation with the algebra (\ref{ALGEBRA}) is
\begin{equation} \label{normes}
E_i =\frac{ S_i}{\sqrt{3}};~~~ E_a = \frac{T_a}{\sqrt{2}}; ~~~E_{ia} = \sqrt{2} G_{ia} .
\end{equation}
Thus, for the generators $S_i$ and $T_a$, which are elements of the $su$(2) and
$su$(3) subalgebras of (\ref{ALGEBRA}), the above expression
simplifies considerably. In particular, as  $S_i$  acts only on the
spin part of the wave function, we apply the usual
Wigner-Eckart theorem for SU(2) to get
\begin{eqnarray}\label{SPIN}
\langle [N_c](\lambda'\mu') Y' I' I'_3; S' S'_3 |S_i|
[N_c](\lambda \mu) Y I I_3; S S_3 \rangle =  \delta_{SS'}\delta_{\lambda \lambda'} \delta_{\mu\mu'} \delta_{YY'} \delta_{II'} \delta_{I_3I_3'}\nonumber \\
  \times \sqrt{C(\mathrm{SU(2)})} \left(\begin{array}{cc|c}
	S   &    1  &  S'   \\
	S_3 &    i  &  S'_3
      \end{array}\right),
   \end{eqnarray}
with $C(\mathrm{SU(2)}) = S(S+1)$.
Similarly, we use the Wigner-Eckart theorem for $T_a$ which is a generator of 
the subgroup SU(3)
\begin{eqnarray}\label{FLAVOR}
\langle [N_c](\lambda'\mu') Y' I' I'_3; S' S'_3 |T_a|
[N_c](\lambda \mu) Y I I_3; S S_3 \rangle =
\delta_{SS'} \delta_{S_3S'_3}\delta_{\lambda \lambda'} \delta_{\mu\mu'}
\nonumber \\
\times
\sum_{\rho = 1,2}
\langle (\lambda'\mu') || T^{(11)} || (\lambda \mu) \rangle_{\rho}
  \left(\begin{array}{cc|c}
	(\lambda \mu)    &  (11)   &   (\lambda'\mu')\\
	YII_3   &  Y^aI^aI^a_{3}  &  Y' I' I'_3
      \end{array}\right)_{\rho},
   \end{eqnarray}
where the reduced matrix element is defined as  \cite{HECHT} 
\begin{eqnarray}\label{REDUCED}
\langle (\lambda \mu) || T^{(11)} || (\lambda \mu) \rangle_{\rho} = \left\{
\begin{array}{cc}
\sqrt{C(\mathrm{SU(3)})}      & \mathrm{for}\ \rho = 1 \\
0 & \mathrm{for}\ \rho = 2 \\
\end{array}\right.
,\end{eqnarray}   
in terms of the eigenvalue of the Casimir operator       
$C(\mathrm{SU(3)}) = \frac{1}{3} g_{\lambda \mu}$ where
\begin{equation}\label{CSU3}
g_{\lambda\mu}= {\lambda}^2+{\mu}^2+\lambda\mu+3\lambda+3\mu.
\end{equation}
The SU(3) CG coefficient factorizes 
into an SU(2)-isospin CG coefficient and an SU(3) isoscalar factor
\cite{DESWART}
\begin{equation}\label{CGSU3}
\left(\begin{array}{cc|c}
	(\lambda \mu)    &  (11)   &   (\lambda'\mu')\\
	YII_3   &  Y^aI^aI^a_{3}  &  Y' I' I'_3
      \end{array}\right)_{\rho} =
\left(\begin{array}{cc|c}
	I   &    I^a  &  I'   \\
	I_3 &    I^a_3  &  I'_3
     \end{array}\right)
 \left(\begin{array}{cc||c}
	(\lambda \mu)    &  (11)   &   (\lambda'\mu')\\
 	 YI   &  Y^aI^a  &  Y' I'
      \end{array}\right)_{\rho}.
 \end{equation}   
The $\rho$ dependence is consistent with Eq. (\ref{GEN}) and reflects the 
multiplicity problem appearing in Eq. (\ref{PROD}) below. 
We shall return to this point in Sec. V.

\section{SU(6) symmetric  wave functions}

Here we consider a wave function which is symmetric in the spin-flavor space. To write
its decomposition into its SU(2)-spin and SU(3)-flavor  parts,
one can use the Kronecker or
inner product of the permutation group S$_n$. The advantage is that one
can treat the permutation symmetry separately in each degree of freedom
\cite{book}.  A basis vector $|[f] Y \rangle $ of an irreducible representation of  S$_n$
is completely defined by the partition $[f]$, and by
a Young tableau $Y$ or its equivalent, an Yamanouchi symbol. In the following 
we do not need to specify the full Young tableau, we only need to know 
the position $p$ of the last particle in each tableau.
In this short-hand notation a symmetric state of $N_c$ quarks is
$|[N_c]1] \rangle$, because $p = 1$. A symmetric spin-flavor wave function can 
be obtained from the product $[f'] \times [f'']$  of spin and flavor states
of symmetries $[f']$ and $[f'']$ respectively, provided  $[f']$ = $[f'']$.

Let us consider a system of $N_c$ quarks having a total spin $S$.
The group SU(2) allows only partitions with maximum two rows, in this case
with  $N_c/2 + S$ boxes in the first row and $N_c/2 - S$
in the second row. So, one has
\begin{equation}\label{FPRIM}
[f'] = [\frac{N_c}{2} + S,\frac{N_c}{2} - S].
\end{equation}
By using the Clebsch-Gordan coefficients of S$_n$ and
their factorization property, described in the Appendix A, one can write
a symmetric state 
of $N_c$ particles with spin $S$ as the linear combination
\begin{eqnarray}\label{FS}
|[N_c ] 1\rangle 
&=& c^{[N_c]}_{11}(S)| [f'] 1 \rangle | [f'] 1 \rangle
  + c^{[N_c]}_{22}(S)| [f'] 2 \rangle | [f'] 2 \rangle,
\end{eqnarray} 
where the coefficients $c^{\mathrm{[N_c]}}_{pp} (p = 1, 2)$ in the right-hand side are
isoscalar factors of the permutation group. 
Their meaning is the following.
The square of the first (second) coefficient is the fraction of Young tableaux
of symmetry $[f']$ having the last 
particle of both states $|[f']p\rangle$ in the first (second) row. That is why they carry the
double index $11$ and $22$ respectively, one index for each state.
Examples of such isoscalar factors can be found in Ref. \cite{ISOSC}.  In the following, the first index refers to the spin part of the wave function and the second index to the flavor part.
The total number of Young
tableaux gives the dimension of the irrep
$[f']$,  so that the sum of squares of the two isoscalar
factors is equal to one.  

In the context of SU(6) $\supset$ SU(2) $\times$ SU(3) there are two alternative forms of each $c^{[N_c]}_{pp}$. They are
derived in Appendix B. The first form is
\begin{eqnarray}\label{SU2}
c^{[N_c]}_{11}(S) & = &  \sqrt{\frac{S[N_c+2(S + 1)]}{N_c(2 S + 1)}}, \nonumber \\
c^{[N_c]}_{22}(S) &  =  & \sqrt{\frac{(S + 1)(N_c - 2 S)}{N_c(2 S + 1)}}.
\end{eqnarray}
These expressions were
obtained by acting with $S_i$ on the spin part of the total wave function and by
calculating the matrix elements of  $S_i$ in two different ways, one
involving the Wigner-Eckart theorem and the other the linear combination
(\ref{FS}).
The coefficients (\ref{SU2}) are precisely the so called ``elements of orthogonal
basis rotation" of Refs. \cite{CCGL} with the identification
$ c^{[N_c]}_{11} = c^{\mathrm{SYM}}_{0-}$ and
$ c^{[N_c]}_{22} = c^{\mathrm{SYM}}_{0+}$.
The other form of the same coefficients, obtained by acting
with $T_a$ on the flavor part of the total wave function is
\begin{eqnarray}\label{SU3}
c^{[N_c]}_{11} (\lambda \mu) & = & \sqrt{\frac{2g_{\lambda\mu}-N_c(\mu-\lambda+3)}{3N_c(\lambda +1)}},      \nonumber \\
c^{[N_c]}_{22} (\lambda \mu) & = & \sqrt{\frac{N_c(6+2\lambda+\mu)-2g_{\lambda\mu}}{3N_c(\lambda+1)}},
\end{eqnarray}
with $g_{\lambda\mu}$ given by Eq. (\ref{CSU3}).  One can use either form, (\ref{SU2}) or (\ref{SU3}),
depending on the  SU(2) or the SU(3) context of the quantity to calculate.
The best is to use the version which leads to simplifications.
One can easily see that the expressions (\ref{SU2}) and (\ref{SU3})
are equivalent to each other. 
By making the replacement $\lambda = 2 S$ and $\mu = N_c/2-S$ in
(\ref{SU3}) one obtains (\ref{SU2}).

The coefficients
$c^{\mathrm{MS}}_{0+}$ and $c^{\mathrm{MS}}_{0-}$ of Refs. \cite{CCGL}, are
also isoscalar factors of the permutation group. They are 
needed to construct a mixed symmetric state from the  inner product
of S$_n$ which generated the symmetric state as well. As shown in Appendix A, they can be obtained from orthogonality relations. The identification is
$ c^{[N_c-1,1]}_{11} = c^{\mathrm{MS}}_{0-}$ and $ c^{[N_c-1,1]}_{22} = c^{\mathrm{MS}}_{0+}$.
There are also the coefficients  $ c^{[N_c-1,1]}_{12} = c^{\mathrm{MS}}_{++}$ and
$ c^{[N_c-1,1]}_{21} = c^{\mathrm{MS}}_{--}$.


\section{Matrix elements of the SU(6) generators}

Besides the standard group theory method of Sec. 2, another method to
calculate the matrix elements of the SU(6) generators
is based on the 
decoupling of the last particle from the 
rest, in each part of the wave function. This is easily done
inasmuch as the row $p$ of the last particle in a Young tableau
is specified.

Let us first consider the spin part. The decoupling is
\begin{equation}\label{statessu(2)}
|S_1, 1/2; S S_3; p\rangle  =
\sum_{m_1,m_2}
 \left(\begin{array}{cc|c}
	S_1  &    1/2  &  S   \\
	m_1  &    m_2  &  S_3 
      \end{array}\right)
|S_1, m_1 \rangle |1/2, m_2 \rangle,
\end{equation}
in terms of an SU(2)-spin CG coefficient
with $S_1 = S - 1/2$ for $p = 1$ and  $S_1 = S + 1/2$ for $p = 2$.

As for the flavor part, a wave function of
symmetry $(\lambda \mu)$ with the last particle in the row $p$
decouples to
\begin{eqnarray}\label{statessu(3)}
\lefteqn{| (\lambda_1 \mu_1) (10); (\lambda \mu)Y I I_{3}; p \rangle  = } 
\nonumber \\ \sum_{Y_1,I_1,I_{1_3},Y_2,I_2,I_{2_3}}
& \left(\begin{array}{cc|c}
	(\lambda_1 \mu_1)    &   (10)  & (\lambda \mu) \\
	 Y_1I_1I_{1_3} &   Y_2I_2I_{2_3} &  YII_{3}
      \end{array}\right) 
|(\lambda_1 \mu_1) Y_1I_1I_{1_3}\rangle
|(10) Y_2I_2I_{2_3} \rangle,
\end{eqnarray}
where $(\lambda_1 \mu_1) = (\lambda - 1, \mu)$ for $p = 1$ and
$(\lambda_1 \mu_1) = (\lambda + 1, \mu - 1)$ for $p = 2$.

Now we use the fact that  $S_i$, $T_a$ and $G_{ia}$ are one-body operators, \emph{i.e.} their general form is
$$O = \sum_{i = 1}^{N_c} O(i).$$
The expectation value of $O$ between symmetric states is equal to $N_c$ times the expectation value of any $O(i)$. Taking $i = N_c$
one has
\begin{equation}\label{ONEB}
\langle O \rangle = N_c \langle O(N_c) \rangle.
\end{equation}
This means that one can  reduce the calculation   of $\langle O \rangle$ to  the calculation of $\langle O(N_c)\rangle$,

To proceed, we recall that the flavor part of $G_{ia}$ is a $T^{(11)}$ tensor in
SU(3). To find out its matrix elements we have to consider the direct product
\begin{eqnarray}\label{PROD}
\lefteqn{(\lambda \mu) \times (11)  =  (\lambda+1, \mu+1)+ (\lambda+2, \mu-1) +
(\lambda \mu)_1 + (\lambda \mu)_2}  \nonumber \\
& & + \, (\lambda-1, \mu+2) + (\lambda-2, \mu+1)
+ (\lambda+1, \mu-2)+ (\lambda-1, \mu-1).
\end{eqnarray}
From the right-hand side, the only terms which give non-vanishing
matrix elements of $G_{ia}$ are $(\lambda' \mu') = (\lambda \mu),
(\lambda+2,\mu-1)$ and $(\lambda-2,\mu+1)$, \emph{i.e.} those with the same
number of boxes, equal to $\lambda + 2\mu$, as on the left-hand side.
In addition, as $G_{ia}$ is a rank 1 tensor in SU(2) it has non-vanishing matrix
elements only for $S'= S, S\pm 1$, \emph{i.e.} again only three distinct
possibilities.
The proper combinations of flavor and spin
parts to get $|[N_c]1
 \rangle$ will be seen in the following subsections.

\subsection{Diagonal matrix elements of $G_{ia}$}

The diagonal matrix element have $(\lambda' \mu') = (\lambda \mu)$
and $S' = S$. 
The first step is to use the relation (\ref{ONEB}) and the
factorization (\ref{FS}) of the spin-flavor wave function into
its spin and flavor parts. This gives
\begin{eqnarray}\label{GIAD}
\lefteqn{\langle [N_c](\lambda \mu) Y' I' I'_3 S S'_3 | G_{ia} |
[N_c](\lambda \mu) Y I I_3 S S_3 \rangle =N_c}  \nonumber \\
& & \times \  \sum_{p = 1,2}
\left({c^{[N_c]}_{pp}(S)}\right)^2 \langle  S_11/2;SS'_3; p\: |s_i(N_c)|\: S_11/2;SS_3; p \rangle \nonumber \\
& & \times \langle (\lambda_1 \mu_1) (10);(\lambda \mu) Y'I'I'_3; p\: |t_a(N_c)|\: (\lambda_1 \mu_1) (10);(\lambda \mu) YII_3; p \rangle,
\end{eqnarray}
where $s_i$ and $t_a$
are the SU(2) and SU(3) generators of the $N_c$-th particle respectively.
The matrix elements of $s_i(N_c)$ between the states (\ref{statessu(2)}) are
\begin{eqnarray}\label{SI}
\lefteqn{\langle S_11/2;SS'_3; p
|s_i(N_c)|S_11/2;SS_3; p \rangle =}
 \nonumber \\
& & \sqrt{\frac{3}{4}}
 \sum_{m_1 m_2 m'_2} 
\left(\begin{array}{cc|c}
	S_1  &    1/2  &  S   \\
	m_1  &    m_2  &  S_3 
      \end{array}\right)
 \left(\begin{array}{cc|c}
	S_1  &    1/2  &  S   \\
	m_1  &    m'_2  &  S'_3 
      \end{array}\right) 
  \left(\begin{array}{cc|c}
	1/2  &    1  &  1/2   \\
	m_2  &    i  &  m'_2   
      \end{array}\right)  \nonumber \\
& &= (-)^{S+S_1-1/2}\sqrt{\frac{3}{2}~(2S+1)}
   \left(\begin{array}{cc|c}
	S   &    1  &   S   \\
	S_3  &   i  &   S'_3   
      \end{array}\right) 
    \left\{\begin{array}{ccc}
	1   &    S  &   S   \\
	S_1  &   1/2  & 1/2
      \end{array}\right\}.
\end{eqnarray}

The matrix elements of the single particle operator $t_a$ between the 
states (\ref{statessu(3)}) are
\begin{eqnarray}\label{TA}
\lefteqn{\langle (\lambda_1 \mu_1) (10); (\lambda \mu) Y' I' I'_3; p
|t_a(N_c)|(\lambda_1 \mu_1) (10); (\lambda \mu) Y I I_3 ; p \rangle
= } \nonumber \\
& & \sqrt{\frac{4}{3}}\sum_{Y_1I_1I_{1_3} Y_2I_2I_{2_3}  Y'_2I'_2I'_{2_3}}\left(\begin{array}{cc|c}
	(\lambda_1 \mu_1)    &  (10)   &   (\lambda \mu)\\
	Y_1I_1I_{1_3}   &  Y_2I_2I_{2_3}  &  YII_3
      \end{array}\right)
     \left(\begin{array}{cc|c}
	(\lambda_1 \mu_1)    &  (10)   &   (\lambda \mu)\\
	Y_1I_1I_{1_3}   &  Y'_2I'_2I'_{2_3}  &  Y' I' I'_3
      \end{array}\right) \nonumber \\
  & &    \times \left(\begin{array}{cc|c}
	(10)    &  (11)   &   (10)\\
	Y_2I_2I_{2_3}   &  Y^aI^aI^a_{3}  &  Y'_2 I'_2 I'_{2_3}
      \end{array}\right)
\nonumber \\
& & = \sqrt{\frac{4}{3}}
  \left(\begin{array}{cc|c}
	I   &    I^a    &   I'   \\
        I_3 &    I^a_3  &   I'_3    
      \end{array}\right)
\sum_{\rho=1,2}\left(\begin{array}{cc||c}
        (\lambda \mu) & (11) &  (\lambda \mu)  \\
	 YI   &    Y^aI^a    &   Y'I'
      \end{array}\right)_ {\rho}\nonumber \\
 & & \times\
 U((\lambda_1,\mu_1)(10)(\lambda \mu)(11);(\lambda \mu)(10))_{\rho}\, ,
\end{eqnarray}
where $U$ are SU(3) Racah coefficients \footnote{To obtain the last equality in (\ref{TA})  we used the identity  \\
$\sum_{\rho=1,2}\langle (\lambda\mu) Y I;(11)Y^aI^a||(\lambda\mu)Y'I'\rangle_{\rho} U((\lambda_1,\mu_1)(10)(\lambda \mu)(11);(\lambda \mu)(10))_{\rho}= \\
\sum_{Y_1I_1 Y_2I_2  Y'_2I'_2}\langle (\lambda_1\mu_1) Y_1 I_1;(10)Y_2I_2||(\lambda\mu)YI\rangle   \langle (10) Y_2I_2;(11)Y^aI^a||(10)Y'_2I'_2\rangle \\
\langle (\lambda_1\mu_1)Y_1I_1;(10)Y'_2I'_2||(\lambda\mu)Y'I'\rangle U(I_1I_2I'I^a;II'_2)$, \\
similar to the relation (12) of Ref. \cite{HECHT}.}.
Note that the sum over $\rho$ is consistent with Eq. (\ref{GEN}) and expresses the fact that the direct product $(\lambda \mu)
\times (11) \rightarrow (\lambda \mu)$ has multiplicity 2 in the
reduction from SU(6) to SU(3), as discussed in Sec. 2.

Introducing (\ref{SI}) and (\ref{TA}) into (\ref{GIAD}) one obtains
\begin{eqnarray}\label{GIADFINAL}
\lefteqn{\langle [N_c](\lambda \mu) Y' I' I'_3 ; S S'_3 | G_{ia} |
[N_c](\lambda \mu) Y I I_3 ; S S_3 \rangle = (-)^{2S} N_c \sqrt{2(2 S + 1)}}
\nonumber \\
  & & \times
  \left(\begin{array}{cc|c}
	S   &   1   & S   \\
	S_3 &   i   & S'_3
      \end{array}\right)           
   \left(\begin{array}{cc|c}
	I   &   I^a   & I'   \\
	I_3 &   I^a_3   & I'_3
      \end{array}\right)
  \sum_{\rho = 1,2}
  \left(\begin{array}{cc||c}
	(\lambda \mu)    &  (11)   &   (\lambda \mu)\\
	Y I   &  Y^a I^a  &  Y' I'
      \end{array}\right)_{\rho} 
      \nonumber \\ 
 & & \times \left[\left({c^{[N_c]}_{22}(S)}\right)^2
    \left\{\begin{array}{ccc}
	S + 1/2   &   1/2   & S   \\
	1       &   S     & 1/2 
      \end{array}\right\}
      U((\lambda+1,\mu-1)(10)(\lambda \mu)(11);(\lambda \mu)(10))_{\rho}\right.
      \nonumber \\
   & &  \left. - \left({c^{[N_c]}_{11}(S)}\right)^2
    \left\{\begin{array}{ccc}
	S - 1/2   &   1/2   & S   \\
	1       &   S     & 1/2 
      \end{array}\right\} 
      U((\lambda-1,\mu)(10)(\lambda \mu)(11);(\lambda \mu)(10))_{\rho}\right],
 \end{eqnarray}
where $c^{[N_c]}_{pp}$ are given by Eqs. (\ref{SU2}) or by the equivalent
form (\ref{SU3}).
 Using the definition of $U$ given in Ref. \cite{HECHT}, Table 4
of the same reference and Table 1 of Ref. \cite{VERGADOS} we have
obtained the following expressions
\begin{equation}
U((\lambda+1,\mu-1)(10)(\lambda \mu)(11);(\lambda \mu)(10))_{\rho=1} =
\frac{\mu - \lambda +3}{4 \sqrt{g_{\lambda \mu}}},
\end{equation}
\begin{equation}
U((\lambda-1,\mu)(10)(\lambda \mu)(11);(\lambda \mu)(10))_{\rho=1} =
\frac{\mu + 2 \lambda +6}{4 \sqrt{g_{\lambda \mu}}},
\end{equation}
\begin{equation}
U((\lambda+1,\mu-1)(10)(\lambda \mu)(11);(\lambda \mu)(10))_{\rho=2} =
\frac{1}{4} \sqrt{
\frac{3 \lambda (\mu+2)(\lambda+\mu+1)(\lambda+\mu+3)}
{\mu (\lambda+2) g_{\lambda \mu}}},
\end{equation}
\begin{equation}
U((\lambda-1,\mu)(10)(\lambda \mu)(11);(\lambda \mu)(10))_{\rho=2} = 
- \frac{1}{4} \sqrt{
\frac{3 (\lambda+2) \mu (\mu+2) (\lambda+\mu+3)}
{{\lambda (\lambda+\mu+1) g_{\lambda \mu}}}},
\end{equation}
where $g_{\lambda \mu}$ is given by the relation (\ref{CSU3}).

\subsection{Off-diagonal matrix elements of $G_{ia}$}

We have applied the procedure of the previous subsection to obtain
the off-diagonal matrix elements of $G_{ia}$ as well. As mentioned above,
there are only two types of non-vanishing matrix elements, those with 
$(\lambda' \mu') = (\lambda+2, \mu-1)$, $S' = S+1$ and those with  
 $(\lambda' \mu')= (\lambda-2, \mu+1)$, $S' = S-1$.
We found that they are given by
\begin{eqnarray}\label{OFF1}
\langle [N_c](\lambda+2, \mu-1) Y' I' I'_3 ; S+1, S'_3 | G_{ia} |
[N_c](\lambda \mu) Y I I_3 ; S S_3 \rangle = (-)^{2S+1} N_c \sqrt{2(2 S + 1)} 
\nonumber \\
  \times
  \left(\begin{array}{cc|c}
	S   &   1   & S+1   \\
	S_3 &   i   & S'_3 
      \end{array}\right)           
   \left(\begin{array}{cc|c}
	I   &   I^a   & I'   \\
	I_3 &   I_3^a   & I'_3
      \end{array}\right)
  \left(\begin{array}{cc||c}
	(\lambda \mu)    &  (11)   &   (\lambda+2, \mu-1)\\
	Y I   &  Y^a I^a  &  Y' I'
      \end{array}\right)
     c^{[N_c]}_{11}(S+1)  c^{[N_c]}_{22}(S) \nonumber \\ 
    \times 
    \left\{\begin{array}{ccc}
	S + 1/2   &   1/2   & S   \\
	1       &   S+1     & 1/2 
      \end{array}\right\} 
      U((\lambda+1,\mu-1)(10)(\lambda+2, \mu-1)(11);(\lambda \mu)(10)),
 \end{eqnarray}
and 
 \begin{eqnarray}\label{OFF2}
\langle [N_c](\lambda-2, \mu+1) Y' I' I'_3 ; S-1,S'_3 | G_{ia} |
[N_c](\lambda \mu) Y I I_3 ; S S_3 \rangle = (-)^{2S} N_c \sqrt{2(2 S + 1)} 
\nonumber \\
  \times
  \left(\begin{array}{cc|c}
	S   &   1   & S-1   \\
	S_3 &   i   & S'_3 
      \end{array}\right)           
   \left(\begin{array}{cc|c}
	I   &   I^a   & I'   \\
	I_3 &   I_3^a   & I'_3
      \end{array}\right)      
  \left(\begin{array}{cc||c}
	(\lambda \mu)    &  (11)   &   (\lambda-2, \mu+1)\\
	Y I   &  Y^a I^a  &  Y' I'
      \end{array}\right) 
     c^{[N_c]}_{22}(S-1)  c^{[N_c]}_{11}(S) \nonumber \\ 
    \times 
    \left\{\begin{array}{ccc}
	S - 1/2   &   1/2   & S   \\
	1       &   S-1     & 1/2 
      \end{array}\right\}
      U((\lambda-1,\mu)(10)(\lambda-2,\mu+1)(11);(\lambda\mu)(10)).
 \end{eqnarray}
Note that the off-diagonal matrix elements do not contain a summation over
$\rho$ because in the right-hand side of the SU(3) product (\ref{PROD}) the terms
$(\lambda+2, \mu-1)$ and $(\lambda-2, \mu+1)$ appear with multiplicity 1.

The above expression require one of the following $U$ coefficients
\begin{equation}
U((\lambda+1,\mu-1)(10)(\lambda+2, \mu-1)(11);(\lambda \mu)(10)) =
\frac{1}{2} \sqrt{\frac{3 (\lambda + \mu + \lambda\mu +1)}
{2 \mu (\lambda + 2)}},
\end{equation}
and 
\begin{equation}
U((\lambda-1,\mu)(10)(\lambda-2, \mu+1)(11);(\lambda \mu)(10)) =
- \frac{1}{2} \sqrt{\frac{3 (\lambda + 1)(\lambda + \mu + 2)}
{2 \lambda (\lambda + \mu + 1)}}.
\end{equation}

As a practical application for $N_c = 3$, the off-diagonal matrix element
are needed to couple $^{4}8$ and $^{2}8$ baryon states, for example.

\section{Isoscalar factors of SU(6) generators for arbitrary $N_c$}

Here we derive analytic formulas for isoscalar factors related to matrix elements 
of SU(6) generators between symmetric states by comparing the definition (\ref{GEN}) 
with results from Sec. 4. In doing this, we have to replace $E_{ia}$ by the 
corresponding generators $S_i$, $T_a$ or $G_{ia}$ according to the 
relations (\ref{normes}). Then from the result (\ref{GIADFINAL}) we obtain the 
isoscalar factor of $G_{ia}$   for $(\lambda' \mu') = (\lambda \mu)$,
$S' = S$ as
\begin{eqnarray}
\lefteqn{\left(\begin{array}{cc||c}
	[N_c]    &  [21^4]   & [N_c]   \\
	(\lambda \mu) S  &  (11) 1  &  (\lambda \mu) S
      \end{array}\right)_{\rho} = N_c (-1)^{2S}
      \sqrt{\frac{4(2S+1)}{C(SU(6))}}}
      \nonumber \\
     & & \times
     \left[ \left({c^{[N_c]}_{22}(S)}\right)^2
    \left\{\begin{array}{ccc}
	S + 1/2   &   1/2   & S   \\
	1       &   S     & 1/2
      \end{array}\right\}
      U((\lambda+1,\mu-1)(10)(\lambda \mu)(11);(\lambda \mu)(10))_{\rho}\right.
      \nonumber \\
    & & - \left.\left({c^{[N_c]}_{11}(S)}\right)^2
    \left\{\begin{array}{ccc}
	S - 1/2   &   1/2   & S   \\
	1       &   S     & 1/2
      \end{array}\right\}
      U((\lambda-1,\mu)(10)(\lambda \mu)(11);(\lambda \mu)(10))_{\rho} \right].
    \end{eqnarray}
Similarly, but using the formula (\ref{OFF2})
we obtain the isoscalar factors for
$(\lambda' \mu') \neq (\lambda \mu)$ $ S' \neq S$. These are
\begin{eqnarray}
\lefteqn{\left(\begin{array}{cc||c}
	[N_c]    &  [21^4]   & [N_c]   \\
	(\lambda \mu) S  &  (11) 1  &  (\lambda+2, \mu-1) S+1
      \end{array}\right) = N_c (-1)^{2S+1}
      \sqrt{\frac{4(2S+1)}{C(SU(6))}}}\nonumber \\
    & & \times \ c^{[N_c]}_{11}(S+1)~c^{[N_c]}_{22}(S)\left\{\begin{array}{ccc}
	S + 1/2   &   1/2   & S   \\
	1       &   S+1     & 1/2
      \end{array}\right\}\nonumber \\
 & & \times \ U((\lambda+1,\mu-1)(10)(\lambda+2,\mu-1)(11);(\lambda \mu)(10)),
\end{eqnarray}
and
\begin{eqnarray}
\lefteqn{\left(\begin{array}{cc||c}
	[N_c]    &  [21^4]   & [N_c]   \\
	(\lambda \mu) S  &  (11) 1  &  (\lambda-2, \mu+1)S-1
      \end{array}\right) = N_c (-1)^{2S}
      \sqrt{\frac{4(2S+1)}{C(SU(6))}}}
      \nonumber \\
      & & \times \ c^{[N_c]}_{11}(S) c^{[N_c]}_{22}(S-1)
    \left\{\begin{array}{ccc}
	S - 1/2   &   1/2   & S   \\
	1       &   S-1     & 1/2
      \end{array}\right\} \nonumber \\
 & & \times \ U((\lambda-1,\mu)(10)(\lambda-2, \mu+1)(11);(\lambda \mu)(10)).
\end{eqnarray}
Replacing $c^{[N_c]}_{pp}$ by definitions (\ref{SU2}) and the $U$ coefficients by their
expressions we have obtained the isoscalar factors for arbitrary  $N_c$ listed in
the first four rows of Table 1.

For completeness, we now return to the generators $S_i$ and $T_a$ which have
only diagonal matrix elements.  
For $S_i$  we use the equivalence between Eq. (\ref{GEN}) and the 
Wigner-Eckart theorem (\ref{SPIN}). This leads to row 5 of Table 1.
For $T_a$ we use the equivalence between Eq. (\ref{GEN}) and 
the Wigner-Eckart theorem (\ref{FLAVOR}). 
The calculation of the isoscalar factors for $\rho = 1$ and $\rho = 2$
gives the results shown in rows 6 and 7 of Table 1.

One can alternatively express the SU(6) isoscalar factors of Table 1 in 
terms of $\lambda$ and $\mu$ by using the identities $\lambda=2S$, 
$\mu=N_c/2-S$ and  $g_{\lambda\mu} = [N_c(N_c+6)+12S(S+1)]/4$.

Before ending this section let us calculate, as an example,  the diagonal
matrix element of $G_{i8}$ by using Eq. (\ref{GEN}). We consider a system of 
$N_c$ quarks with spin $S$, isospin $I$ and strangeness $\mathcal{S}$ defined by 
$Y=N_c/3 + \mathcal{S}$. In this case Eq. (\ref{GEN}) becomes
\begin{eqnarray}\label{EXAMPLE}
\langle [N_c](\lambda\mu) Y'I'I'_3 S S'_3 | G_{i8} |
[N_c](\lambda \mu) Y I I_3 S S_3 \rangle =  \delta_{YY'}\delta_{II'}\delta_{I_3I'_3}\sqrt{\frac{C(\mathrm{SU(6)})}{2}}
 \left(\begin{array}{cc|c}
	S   &    1   & S   \\
	S_3  &   i   & S'_3
  \end{array}\right)
   \nonumber \\
 \times     \left(\begin{array}{cc|c}
	I   &   0  & I'   \\
	I_3 &   0   & I'_3
   \end{array}\right)    \sum_{\rho = 1,2}
 \left(\begin{array}{cc||c}
	(\lambda \mu)    &  (11)   &   (\lambda \mu)\\
	Y I   &  00  &  Y' I'
      \end{array}\right)_{\rho}
\left(\begin{array}{cc||c}
	[N_c]    &  [21^4]   & [N_c]   \\
	(\lambda \mu) S  &  (11)1  &  (\lambda \mu) S
      \end{array}\right)_{\rho} , 
   \end{eqnarray}
Using Table 4 of Ref. \cite{HECHT} and our Table 1 for the isoscalar 
factors of SU(3) and SU(6) respectively we have obtained
\begin{eqnarray}
\langle [N_c](\lambda\mu) Y'I' I'_3 S S'_3 | G_{i8} |
[N_c](\lambda \mu) Y I I_3 S S_3 \rangle =
 \frac{\delta_{YY'}\delta_{II'}\delta_{I_3I'_3}}{ 4 \sqrt{3S(S+1)}}
  \left(\begin{array}{cc|c}
	S   &    1   & S   \\
	S_3  &   i   & S'_3
  \end{array}\right) \nonumber \\
    \times \,
   [3 I (I+1) - S (S+1) - \frac{3}{4} \mathcal{S} (\mathcal{S} - 2 )].
   \end{eqnarray}
With $\mathcal{S} = -N_s$, where $N_s$ is the number of strange quarks,
we can recover the relation 
\begin{equation}
S_i G_{i8} = \frac{1}{4 \sqrt{3}}
 [3 I (I+1) - S (S+1) -\frac{3}{4} N_s ( N_s + 2 )],
\end{equation}
used in our previous work \cite{MS1}. (Ref.  \cite{MS1}  contains a typographic
error. In the denominator of Eq. (13) one should read $\sqrt{3}$ instead of
$\sqrt{2}$.)

\section{Back to isoscalar factors of SU(4) generators}

In the case of SU(4) $\supset$ SU(2) $\times$ SU(2) the analogue of Eq. (\ref{GEN}) is \cite{HP}
\begin{eqnarray}\label{GENsu4}
\langle [N_c] I' I'_3 S' S'_3 | E_{ia} |
[N_c] I I_3 S S_3 \rangle = \sqrt{C(\mathrm{SU(4)})}   \nonumber \\
\times
\left(\begin{array}{cc||c}
	[N_c]    &  [21^2]   & [N_c]   \\
	I S  &   I^a S^i  &  I'S'
      \end{array}\right) 
  \left(\begin{array}{cc|c}
	S   &    S^i   & S'   \\
	S_3  &   S^i_3   & S'_3
  \end{array}\right)
     \left(\begin{array}{cc|c}
	I   &   I^a   & I'   \\
	I_3 &   I^a_3   & I'_3
   \end{array}\right),
   \end{eqnarray}
where $C(\mathrm{SU(4)})=[3N_c(N_c+4)]/8$ is the eigenvalue of the SU(4) Casimir operator. Note that for a symmetric state one has $I=S$.
We recall that the SU(4) algebra is
\begin{eqnarray}\label{ALGEBRASU4}
[S_i,S_j] & = & i \varepsilon_{ijk} S_k,
\ \ \ [T_a,T_b]  =  i \varepsilon_{abc} T_c, \ \ \ [S_i,T_a]  =  0, \nonumber \\
\lbrack S_i,G_{ia}\rbrack & = & i\varepsilon_{ijk}G_{ka}, \ \ \ [T_a,G_{ib}]=i\varepsilon_{abc}G_{ic}, \nonumber \\
\lbrack G_{ia},G_{jb}\rbrack & =  & \fr{i}{4} \delta_{ij} \varepsilon_{abc} T_c
+\fr{i}{2} \delta_{ab}\varepsilon_{ijk}S_k.
\end{eqnarray}
The tensor operators $E_{ia}$ are related to $S_i$, $T_a$ and $G_{ia}$ $(i=1,2,3;\ a=1,2,3)$ by
\begin{equation} \label{normes2}
E_i =\frac{S_i}{\sqrt{2}};~~~ E_a = \frac{T_a}{\sqrt{2}}; ~~~E_{ia} = \sqrt{2} G_{ia}.
\end{equation}
In Eq. (\ref{GENsu4}) they are identified by $I^aS^i=01,10$ and 11 respectively. Now we want to obtain the SU(4) isoscalar factors as particular cases of the SU(6) results with $Y^a=0$. In SU(4) the hypercharge of a system of $N_c$ quarks takes the value $Y=N_c/3$.
By comparing (\ref{GEN}) and (\ref{GENsu4}) we obtained the relation
\begin{eqnarray}\label{isosu4}
\lefteqn{\left(\begin{array}{cc||c}
	[N_c]    &  [21^2]   & [N_c]   \\
	I S  &   I^a S^i  &  I'S'
      \end{array}\right)
 = r^{I^aS^i} \sqrt{\frac{C(\mathrm{SU(6))}}{C(\mathrm{SU(4))}}}} \nonumber \\
 & & \times \sum_{\rho = 1,2}\left(\begin{array}{cc||c}
	(\lambda \mu)    &  (\lambda^a\mu^a)   &   (\lambda' \mu')\\
	\frac{N_c}{3} I   &  0 I^a  &  \frac{N_c}{3} I'
      \end{array}\right)_{\rho}
\left(\begin{array}{cc||c}
	[N_c]    &  [21^4]   & [N_c]   \\
	(\lambda \mu) S  &  (\lambda^a\mu^a) S^i  &  (\lambda' \mu') S'
      \end{array}\right)_{\rho}\, ,
\end{eqnarray}
where
\begin{eqnarray}
r^{I^aS^i}=\left\{
\begin{array}{cc} \sqrt{\frac{3}{2}} & \mathrm{for}\ I^aS^i=01 \\
1 &\mathrm{for}\ I^aS^i=10 \\
1 &\mathrm{for}\ I^aS^i=11
\end{array}\right.
,\end{eqnarray}
due to (\ref{normes}), (\ref{GENsu4}) and (\ref{normes2}). In Eq. (\ref{isosu4}) 
we have to make the replacement
\begin{equation}
\lambda = 2I,\ \mu=\frac{N_c}{2}-I;\
\lambda' = 2I',\  \mu'=\frac{N_c}{2}-I',
\end{equation}
and take
\begin{eqnarray}
(\lambda^a\mu^a)=\left\{
\begin{array}{cc}
(00) & \mathrm{for}\ I^a=0 \\
(11) & \mathrm{for}\ I^a=1 \\
\end{array}\right.
.\end{eqnarray}
In this way we recovered the SU(4) isoscalar factors presented in  Table A4.2 of Ref. \cite{HP} up to a phase factor.
By introducing these isoscalar factors into the matrix elements (\ref{GENsu4}) we obtained the expressions given in Eqs. (A1-A3) of  Ref. \cite{CCGL}.
\section{The mass operator}

As mentioned in the introduction, in calculating the mass spectrum of 
baryons belonging to $[70]$-plets, the
general procedure is to decompose the generators of a system of $N_c$
quarks representing a large $N_c$ baryon, into sums of core and excited
quark generators. In order to apply these separate parts one also has
to decouple the excited quark from the core. As an essential ingredient, 
the spin-flavor part of the core wave function is totally symmetric. 
Then the matrix elements of the core operators can be calculated by 
using Table I where one has to replace $N_c$ by $N_c-1$.
  
When the SU(3) symmetry is exact, the mass operator of an excited state 
can be written as the linear combination 
\begin{equation}
\label{massoperator}
M = \sum_{i} c_i O_i ,
\end{equation} 
where $c_i$ are unknown coefficients which parametrize the QCD dynamics
and the operators $O_i$ are of type 
\begin{equation}\label{OLFS}
O_i = \frac{1}{N^{n-1}_c} O^{(k)}_{\ell} \cdot O^{(k)}_{SF}
\end{equation}
where  $O^{(k)}_{\ell}$ is a $k$-rank tensor in SO(3) and  $O^{(k)}_{SF}$
a $k$-rank tensor in SU(2), but scalar in SU(3)-flavor.
This implies that $O_i$ is a combination 
of SO(3) generators $\ell_i$ and of SU(6) generators. 
Additional operators are needed when SU(3) is broken.
The values of the coefficients $c_i$  are found by  
a numerical fit to data.

An essential step is to find all linearly independent operators 
contributing to a given order $\mathcal{O}(1/N_c)$. This problem
was extensively discussed for example in Ref. \cite{CCGL} for 
excited baryons belonging to the $[{\bf 70},1^-]$ multiplet. For
$[{\bf 70},\ell^+]$ the problem is more complicated because the 
core can also be excited.  
However the practice on the $[{\bf 70},1^-]$ multiplet
showed that some operators are dominant \cite{CCGL,SGS},
which is expected to remain  valid for the $[{\bf 70},\ell^+]$
multiplet as well \cite{MS2}.
These are 
$O_1 = N_c \ \1 $, $O_2 = \ell_q^i s^i$, 
$O_3 = \frac{3}{N_c}\ell^{(2)ij}_{q}g^{ia}G_c^{ja}$, 
$O_4 = \frac{4}{N_c+1} \ell^i t^a G_c^{ia}$ and
$O_5 = \frac{1}{N_c}(S_c^iS_c^i+s^iS_c^i)$.   
Among them, the operators $O_3$ and $O_4$ contain the core generator
$G_c^{ia}$. Table I provides the necesary matrix elements for
this generator when the study is extended to strange baryons. 
 Applications to baryons belonging to the $[{\bf 70},\ell^+]$
multiplet are presented elsewhere \cite{MS4} where problems related
to the corresponding description of the wave functions are also discussed. 

\section{Summary}
We have derived general explicit formulas for specific matrix elements
of the SU(6) generators as a function of $N_c$.
They refer to spin-flavor symmetric states $|[N_c] 1 \rangle$.
They can be applied, for example, to the study of baryon excited states
belonging to the $[{\bf 70},\ell]$ multiplet.
In that case the system of $N_c$ quarks describing the baryon is divided into
an excited quark and a core, which is the remaining of a baryon state
after an excited quark has been removed. The core is always
described by a symmetric spin-flavor state. 
In using
Table I for the core operators one must replace $N_c$ by $N_c-1$.
The results of Table I can equally be applied to the study of pentaquarks
\cite{PS}

Putting $N_c=3$ in Table I we could, for example, reproduce the values of Refs. \cite{CCM} and \cite{SCHULKE} up to a phase.
To further check the validity of our results we have calculated the matrix elements
of the operator $O_3$ \cite{SGS}, which contains $G_{ia}$.
We have recovered the expressions of $O_3$ needed to calculate masses
of the strange and nonstrange baryons belonging to the $[{\bf 70},1^-]$
multiplet.


\appendix
\section{}
Here we shortly recall the definition of isoscalar factors of the
permutation group S$_n$. Let us denote a basis vector in the
invariant subspace of the irrep
$[f]$ of S$_n$ by $|[f] Y \rangle $, where $Y$ is the
corresponding Young tableau or Yamanouchi symbol. A basis vector obtained
from the inner product of two irreps $[f']$ and $[f'']$ is defined by
the sum over  products of basis vectors of $[f']$ and $[f'']$ as
\begin{equation}
|[f] Y \rangle = \sum_{Y'Y''}
S([f']Y' [f'']Y'' | [f]Y ) |[f']Y'\rangle |[f'']Y'' \rangle,
\end{equation}
where $S([f']Y' [f'']Y'' | [f]Y )$ are  Clebsch-Gordan (CG) coefficients
of S$_n$.
Any CG coefficient can be factorized into an isoscalar factor, here called
$K$ matrix \cite{book}, and a CG coefficient of S$_{n-1}$. To apply the
factorization property it is necessary to specify the row $p$ of the 
$n$-th particle and the row $q$ of the $(n-1)$-th particle. The remaining
particles are distributed in a Young tableau denoted by $y$.
Then the isoscalar factor $K$ associated to a given CG of S$_n$ is defined as
\begin{equation}
S([f']p'q'y' [f'']p''q''y'' | [f]pqy ) =
K([f']p'[f'']p''|[f]p) S([f'_{p'}]q'y' [f''_{p''}]q''y'' | [f_p]qy ),
\end{equation}
where  the right-hand side contains a CG coefficient
of S$_{n-1}$ containing $[f_p]$, $[f'_{p'}]$ and $[f''_{p''}]$ which are the partitions obtained from $[f]$ after
the removal of the $n$-th particle.
The $K$ matrix obeys the following orthogonality relations
\begin{eqnarray}
\sum_{p'p''}  K([f']p'[f'']p''|[f]p) K([f']p'[f'']p''|[f_1]p_1) & = &\delta_{f f_1}
\delta{p p_1}, \label{K1}\\
\sum_{fp}  K([f']p'[f'']p''|[f]p) K([f']p'_1[f'']p''_1|[f]p) & = &
\delta_{p'p'_1} \delta_{p'' p''_1} \label{K2}.
\end{eqnarray}
The isoscalar factors used to construct the spin-flavor symmetric state
(\ref{FS}) are
\begin{eqnarray}\label{SYM}
c^{[N_c]}_{11} & = & K([f']1[f']1|[N_c]1),\nonumber \\
c^{[N_c]}_{22} & = & K([f']2[f']2|[N_c]1),
\end{eqnarray}
with $[f'] = [N_c/2+S,N_c/2-S]$.
The isoscalar factors needed to construct the state of mixed
symmetry $[N_c-1,1]$ from the same inner product are
\begin{eqnarray}\label{MS}
c^{[N_c-1,1]}_{11}  & = & K([f']1[f']1|[N_c-1,1]2),\nonumber \\
c^{[N_c-1,1]}_{22}  & = & K([f']2[f']2|[N_c-1,1]2).
\end{eqnarray}
The above coefficients and the orthogonality
relation (\ref{K2}) give 
\begin{eqnarray}
c_{11}^{[N_c-1,1]} & = & -c_{22}^{[N_c]}, \nonumber \\
c_{22}^{[N_c-1,1]} & = & c_{11}^{[N_c]}.
\end{eqnarray}

When the last particle is located in different rows in the flavor and
spin parts the needed coefficients are
\begin{eqnarray}
c^{[N_c-1,1]}_{12}  & = & K([f']1[f']2|[N_c-1,1]2) = 1,\nonumber \\
c^{[N_c-1,1]}_{21}  & = & K([f']2[f']1|[N_c-1,1]2) = 1,
\end{eqnarray}
which are identical because of the symmetry properties of $K$.
For $N_c \leq $ 6 one can check the above identification with
the isoscalar factors given in Ref. \cite{ISOSC}.
The identification of the so called ``elements of orthogonal basis
rotation" of Ref. \cite{CCGL} with the above isoscalar factors of S$_n$ is given in Sec. III.
The expressions of these coefficients
are derived in the following appendix for arbitrary $N_c$.
\section{}
Here we derive the expressions of the coefficients
$\left(c_{pp}^{[N_c]}\right)^2$ ($p = 1,2$),
defined in the context of SU(6) $\supset$ SU(2) $\times$  SU(3) as  isoscalar factors of S$_n$.
 To get (\ref{SU2}) we write the matrix elements of the generators $S_i$
in two different ways. One is to use the Wigner-Eckart theorem (\ref{SPIN}).
The other is  to calculate the matrix
elements of $S_i$ by using (\ref{FS}), (\ref{ONEB}) and (\ref{SI}). By comparing the two expressions we obtain the equality
\begin{eqnarray}\label{B1}
\sqrt{S(S+1)}  & = &  (-)^{2S} N_c \sqrt{\frac{3}{2}}\sqrt{2S+1}
\left[\left(c_{22}^{[N_c]}\right)^2 \left\{\begin{array}{ccc} 1 & S & S \\ S+1/2 & 1/2 & 1/2 \end{array}\right\}\right. \nonumber \\ & & - \left.\left(c_{11}^{[N_c]}\right)^2 \left\{\begin{array}{ccc} 1 & S & S \\ S-1/2 & 1/2 & 1/2 \end{array}\right\}\right],
\end{eqnarray}
which is an equation for the unknown quantities 
The other equation is the normalization relation (\ref{K1})
\begin{equation}\label{normaliz}
\left(c_{11}^{[N_c]}\right)^2 +\left(c_{22}^{[N_c]}\right)^2 = 1.
\end{equation}
We found
\begin{eqnarray}
c^{[N_c]}_{11}(S) & = & \sqrt{\frac{S[N_c+2(S + 1)]}{N_c(2 S + 1)}}, \nonumber \\
c^{[N_c]}_{22}(S) & = & \sqrt{\frac{(S + 1)(N_c - 2 S)}{N_c(2 S + 1)}}.
\end{eqnarray}
which are  the relations (\ref{SU2}). This phase convention is consistent
with Ref. \cite{ISOSC}.
Similarly, to get (\ref{SU3}) we calculate the matrix elements of the generators
$T_a$ from (\ref{FS}), (\ref{ONEB}) and (\ref{TA}) and compare to the
Wigner-Eckart theorem (\ref{FLAVOR}).
This leads to the equation
\begin{eqnarray}\label{B4}
\sqrt{\frac{g_{\lambda\mu}}{3}} &  =  &
  \frac{2}{\sqrt{3}} N_c \left[\left(c_{22}^{[N_c]}\right)^2 U((\lambda+1,\mu-1)(10)(\lambda\mu)(11);(\lambda\mu)(10))_{\rho=1}\right. \nonumber \\
 & & + \left.\left(c_{11}^{[N_c]}\right)^2  U((\lambda-1,\mu)(10)(\lambda\mu)(11);(\lambda\mu)(10))_{\rho=1}\right],
\end{eqnarray}
which together with the normalization condition (\ref{normaliz}) give
\begin{eqnarray}
c^{[N_c]}_{11} (\lambda \mu) & = & \sqrt{\frac{2g_{\lambda\mu}-N_c(\mu-\lambda+3)}{3N_c(\lambda +1)}},      \nonumber \\
c^{[N_c]}_{22} (\lambda \mu) & = & \sqrt{\frac{N_c(6+2\lambda+\mu)-2g_{\lambda\mu}}{3N_c(\lambda+1)}}.
\end{eqnarray}
\emph{i.e.} the relations (\ref{SU3}).
In addition, we found that the following identity holds for $\rho = 2$
\begin{eqnarray}\label{B5}
0 & = &
  \frac{2}{\sqrt{3}} N_c
  \left[\left(c_{22}^{[N_c]}\right)^2
  U((\lambda+1,\mu-1)(10)(\lambda\mu)(11);(\lambda\mu)(10))_{\rho=2}\right. \nonumber \\
 & & + \left.\left(c_{11}^{[N_c]}\right)^2 
 U((\lambda-1,\mu)(10)(\lambda\mu)(11);(\lambda\mu)(10))_{\rho=2}\right].
\end{eqnarray}
 This  cancellation is consistent with the definition of the matrix elements
 of the SU(3) generators Eqs. (\ref{FLAVOR}), (\ref{REDUCED}) and it is an 
 important check of our results.


\begin{sidewaystable}
\renewcommand{\arraystretch}{1.5}
\begin{tabular}{l|c|c|l}
\hline
\hline

$(\lambda_1\mu_1)S_1$ \hspace{0.5cm} & \hspace{0.5cm}$(\lambda_2\mu_2)S_2$ \hspace{0.5cm} &\hspace{0.5cm}$\rho$\hspace{0.5cm} & \hspace{0.5cm}$\left(\begin{array}{cc||c}                                         [N_c]  &  [21^4]  &  [N_c] \\
                           (\lambda_1\mu_1)S_1 & (\lambda_2\mu_2)S_2 & (\lambda\mu)S
                                      \end{array}\right)_\rho$  \\
\vspace{-0.8cm} &  &   & \\
\hline
$(\lambda + 2,\mu - 1)S+1$\hspace{0.5cm} & \hspace{0.0cm}$(11)1$ & $/$ &\hspace{0.5cm}$-\sqrt{\frac{3}{2}}\sqrt{\frac{2S+3}{2S+1}}\sqrt{\frac{(N_c-2S)(N_c+2S+6)}{5N_c(N_c+6)}}$ \\
$(\lambda\mu)S$  & \hspace{0.0cm}$(11)1$ & 1 & \hspace{0.5cm}$4(N_c+3)\sqrt{\frac{2S(S+1)}{5N_c(N_c+6)[N_c(N_c+6)+12S(S+1)]}}$ \\
$(\lambda\mu)S$  &\hspace{0.0cm}$(11)1$ & 2 & \hspace{0.5cm}$-\sqrt{\frac{3}{2}}\sqrt{\frac{(N_c-2S)(N_c+4-2S)(N_c+2+2S)(N_c+6+2S)}{5N_c(N_c+6)[N_c(N_c+6)+12S(S+1)]}}$ \\
$(\lambda - 2,\mu + 1)S-1$  & \hspace{0.0cm}$(11)1$  & $/$ & \hspace{0.5cm}$-\sqrt{\frac{3}{2}}\sqrt{\frac{2S-1}{2S+1}}\sqrt{\frac{(N_c+4-2S)(N_c+2+2S)}{5N_c(N_c+6)}}$ \\
$(\lambda\mu)S$  & \hspace{0.0cm}$(00)1$ & $/$  & \hspace{0.5cm}$\sqrt{\frac{4S(S+1)}{5N_c(N_c+6)}}$ \\
$(\lambda\mu)S$  & \hspace{0.0cm}$(11)0$ & $1$  & \hspace{0.5cm}$\sqrt{\frac{N_c(N_c+6)+12S(S+1)}{10N_c(N_c+6)}}$ \\
$(\lambda\mu)S$  & \hspace{0.0cm}$(11)0$ & $2$  & \hspace{0.5cm}0 \\
\hline
\hline
\end{tabular}
\caption{Isoscalar factors SU(6) for $[N_c] \times [21^4] \rightarrow [N_c]$ defined by Eq. (\ref{GEN}).}
\end{sidewaystable}


\begin{thebibliography}{11}
\bibitem{HOOFT} G. 't Hooft, Nucl. Phys. {\bf 72} (1974) 461.

\bibitem{WITTEN} E. Witten, Nucl. Phys. {\bf B160} (1979) 57.

\bibitem{DM} R. Dashen and A. V. Manohar, Phys. Lett. {\bf B315} (1993) 425;
ibid {\bf B315} (1993) 438.


\bibitem{DJM94}
R.~Dashen, E.~Jenkins, and A.~V.~Manohar,
Phys.~Rev.  {\bf D49} (1994) 4713.
\bibitem{DJM95}
R. Dashen, E. Jenkins, and A. V. Manohar,
Phys. Rev. {\bf D51} (1995) 3697.
\bibitem{CGO94}
C.~D.~Carone, H.~Georgi and S.~Osofsky,
Phys.~Lett. {\bf B322} (1994) 227.\\
M.~A.~Luty and J.~March-Russell,
Nucl.~Phys. {\bf B426} (1994) 71.\\
M.~A.~Luty, J.~March-Russell and  M.~White,
Phys.~Rev.  {\bf D51} (1995)  2332.
\bibitem{Jenk1}
E.~Jenkins, Phys.~Lett.  {\bf B315} (1993) 441.
\bibitem{JL95}
E.~Jenkins and R.~F.~Lebed,
Phys.~Rev.  {\bf D52} (1995) 282.
\bibitem{DDJM96}
J.~Dai, R.~Dashen, E.~Jenkins, and A.~V.~Manohar,
Phys.~Rev.   {\bf D53} (1996) 273.

\bibitem{CGKM} C. D. Carone, H. Georgi, L. Kaplan and D. Morin,
 Phys. Rev. {\bf D50} (1994) 5793.

\bibitem{Goi97}
J.~L. Goity, Phys. Lett.  {\bf B414} (1997) 140.

\bibitem{PY1} D. Pirjol and T.~M. Yan,
 Phys. Rev.  {\bf D57}  (1998) 1449.

\bibitem{PY2}  D. Pirjol and T.~M. Yan,  Phys. Rev. {\bf D57} (1998) 5434.

\bibitem{CCGL}
C.~E. Carlson, C.~D. Carone, J.~L. Goity and R.~F. Lebed,
 Phys. Lett. {\bf B438}  (1998) 327;
 Phys. Rev. {\bf D59}  (1999) 114008.


\bibitem{CaCa98}
C.~E.~Carlson and C.~D.~Carone,
 Phys.\ Lett.\  {\bf  B441} (1998) 363;
 Phys.\ Rev.\  {\bf  D58} (1998) 053005.

\bibitem{BCCG} Z. A. Baccouche, C. K. Chow, T. D. Cohen and B. A. Gelman,
 Nucl.\ Phys.\  {\bf  A696}  (2001) 638.
 
\bibitem{SCHAT} C. L. Schat, 
\emph{Tempe 2002, Phenomenology of large N(c) QCD}, (2002) 189-198 [arXiv:hep-ph/0204044].

\bibitem{Pirjol:2003ye}
D.~Pirjol and C.~Schat, Phys. Rev. {\bf  D67} (2003) 096009.

\bibitem{cohen1}
T. D. Cohen, D. C. Dakin, A. Nellore and R. F. Lebed,
Phys. Rev. {\bf D69} (2004) 056001.

\bibitem{SGS}
C.~L. Schat, J.~L. Goity and N.~N. Scoccola,  Phys. Rev. Lett.
{\bf 88} (2002) 102002;
J.~L.~Goity, C.~L.~Schat and N.~N.~Scoccola,
 Phys.\ Rev.\  {\bf  D66} (2002) 114014.
 

\bibitem{CC00}
C. E. Carlson and C. D. Carone,
Phys. Lett. {\bf B484} (2000) 260.

\bibitem{GSS}
J.~L. Goity,   C.~L.~Schat and N.~N.~Scoccola,
 Phys. Lett. {\bf B564} (2003) 83.

\bibitem{MS2} N. Matagne and Fl. Stancu, Phys. Lett {\bf B631} (2005) 7.

\bibitem{MS1} N. Matagne and Fl. Stancu,  Phys. Rev. {\bf D71} (2005) 014010.

\bibitem{GOITY05} J. L. Goity, 
\emph{Trento 2004, Large N(c) QCD}, (2004) 211-222 [arXiv:hep-ph/0504121].

\bibitem{CL} T. D. Cohen and R. F. Lebed, Phys. Rev. {\bf D70} (2004) 096015.

\bibitem{HP} K. T. Hecht and S. C. Pang, J. Math. Phys. {\bf 10} (1969) 1571.

\bibitem{CCM} J. C. Carter, J. J. Coyne and S. Meshkov, Phys. Rev. Lett.
{\bf 14} (1965) 523.

\bibitem{SCHULKE} L. Sch\"ulke,  Z. Phys. {\bf 183} (1965) 424.

\bibitem{CC} J. C. Carter and J. J. Coyne J. Math. Phys. {\bf 10}  (1969) 1204.


\bibitem{HECHT} K. T. Hecht, Nucl. Phys. {\bf 62} (1965) 1.

\bibitem{DESWART} J. J. De Swart, Rev. Mod. Phys. {\bf 35} (1963) 916.

\bibitem{ELLIOTT} J. P. Elliott, 
Proc. Roy. Soc. (London) {\bf A245} (1958) 128; ibid. {\bf A245} (1958) 562.


\bibitem{book} Fl. Stancu, {\it Group Theory in Subnuclear Physics},
Clarendon Press, Oxford, (1996) Ch. 4.


\bibitem{ISOSC} Fl. Stancu and S. Pepin, Few-Body Systems, {\bf 26} (1999) 113.

\bibitem{VERGADOS} J. D. Vergados, Nucl. Phys. {\bf A111} (1968) 681.


\bibitem{MS4} N. Matagne and Fl. Stancu, [arXiv:hep-ph/0604122].

\bibitem{PS} D. Pirjol and C. Schat, [arXiv:hep-ph/0603127]

\end{thebibliography}
\end{document}